\documentclass[prd,12pt,tightenlines,nofootinbib]{revtex4}

\usepackage[english]{babel}
\usepackage{graphicx}
\usepackage{psfrag}
\usepackage{amsmath}
\usepackage{amssymb}
\usepackage{bbm}
\usepackage{slashed}
\usepackage{verbatim}
\usepackage[hypertex]{hyperref}

\newcommand{\be}{\begin{equation}}
\newcommand{\ee}{\end{equation}}
\newcommand{\bea}{\begin{eqnarray}}
\newcommand{\eea}{\end{eqnarray}}
\newcommand{\bean}{\begin{eqnarray*}}
\newcommand{\eean}{\end{eqnarray*}}

\renewcommand{\b}{\langle}
\newcommand{\ket}{\rangle}

\newcommand{\irm}{{\rm i}}
\newcommand{\e}{{\rm e}}

\renewcommand{\d}{{\rm d}}
\newcommand{\cl}[1]{{\mathcal #1}}

\newcommand{\pa}{\partial}

\renewcommand{\v}[1]{\vec{#1}}

\newcommand{\ds}{\displaystyle}

\newcommand{\diff}[2]{\frac{{\rm d}#1}{{\rm d}#2}}

\newcommand{\bZ}{\mathbb{Z}}
\newcommand{\bN}{\mathbb{N}}
\newcommand{\bC}{\mathbb{C}}
\newcommand{\bR}{\mathbb{R}}

\newcommand{\clH}{\cl{H}}

\newcommand{\clD}{\cl{D}}

\newcommand{\clF}{\cl{F}}

\newcommand{\clC}{\cl{C}}

\newcommand{\eq}[1]{(\ref{#1})}
\renewcommand{\sec}[1]{sec.\ \ref{#1}}

\newcommand{\twomatrix}[4]
{\left(\begin{array}{rc}#1 & #2 \\ #3 & #4\end{array}\right)}

\newcommand{\qed}{\nobreak \ifvmode \relax \else
      \ifdim\lastskip< 1 em \hskip-\lastskip
      \hskip1.0em plus0em minus0.5em \fi \nobreak
      \vrule height0.75em width0.75em depth0 em\fi}

\newcommand{\xib}{\overline{\xib}}

\newcommand{\SLC}{\mathrm{SL(2,\bC)}}

\newcommand{\Ct}{\tilde{C}}
\newcommand{\Nt}{\tilde{N}}
\newcommand{\Oh}{\hat{O}}

\newcommand{\half}{\frac{1}{2}}
\newcommand{\zb}{\overline{z}}
\newcommand{\vbar}{\overline{v}}

\begin{document}

\title{Unitary irreducible representations of $\SLC$ \\ in discrete and continuous $\mathrm{SU(1,1)}$ bases}

\author{Florian Conrady}
\email{fconrady@perimeterinstitute.ca}
\affiliation{Perimeter Institute for Theoretical Physics, Waterloo, Ontario, Canada}
\author{Jeff Hnybida}
\email{jhnybida@perimeterinstitute.ca}
\affiliation{Perimeter Institute for Theoretical Physics, Waterloo, Ontario, Canada}
\affiliation{Department of Physics, University of Waterloo, Waterloo, Ontario, Canada}

\begin{abstract}
We derive the matrix elements of generators of unitary irreducible representations of $\SLC$
with respect to basis states arising from a decomposition into irreducible representations of SU(1,1).
This is done with regard to a discrete basis diagonalized by $J^3$ and a continuous basis diagonalized by $K^1$,
and for both the discrete and continuous series of SU(1,1).
For completeness we also treat the more conventional SU(2) decomposition as a fifth case.
The derivation proceeds in a functional / differential framework
and exploits the fact that state functions and differential operators have a similar structure
in all five cases. The states are defined explicitly and related to SU(1,1) and SU(2) matrix elements.
\end{abstract}

\maketitle

\section{Introduction}
\label{introduction}

Unitary irreducible representations of $\SLC$ play a central role in
loop and spin foam quantum gravity \cite{Rovellibook,Thiemannbook}.
States in these representations are used as quanta of a field theory that generates
spacetime \cite{ReisenbergerRovelliFeynman1,ReisenbergerRovelliFeynman2,Oritigroupfieldtheoryapproach}.
More precisely, spacetime appears in the form of cell complexes
that are dual to Feynman diagrams.
The states of the unitary irreducible representations of the Lorentz group describe 2--cells and propagate along the lines
of diagrams. They interact at vertices to form 4--cells and thereby give rise to the cell complex.
The quantum numbers are spins and encode the geometry of 2--cells.
A given assignment of spins to all 2--cells is a \textit{spin foam}.
The perturbative expansion results in a sum over cell complexes (Feynman diagrams) and
geometries (quantum numbers) and can be seen as a version of Wheeler's spacetime foam.

Recent years saw considerable progress in this approach
to quantum gravity. It was understood how states have to be constrained to reflect
the geometry of a 2--cell correctly \cite{BarrettCrane,EPRL,FK,Livine_Consistently}. The link between quantum states and
classical geometry  was greatly clarified through the use of coherent states  \cite{Livine_New}.
Remarkably, the simplest possible interaction between these quanta leads to amplitudes
that are closely related to Regge geometry \cite{CFpathrep,CFsemiclassical,BarrettasymptoticsEuclidean,BarrettasymptoticsLorentzian,CFquantumgeometry}.

In order to encode the full structure of a Lorentzian geometry, one requires both
spacelike and timelike 2--cells. As was shown by the authors recently, the latter are implemented
by certain irreps in the SU(1,1) decomposition of $\SLC$ irreps \cite{CHtimelike,Ccoherent}.
The constraints on these irreps were obtained by constructing coherent states that mimic the properties
of classical timelike 2--cells. It was required, in particular, that expectation values of $\SLC$
generators behave like classical bivectors of a timelike triangle.
It turned out that the usual eigenstates of $J^3$ are not suitable to build such coherent states.
Instead we had to use eigenstates of the generator $K^1$ \cite{LindbladNagel} and compute
the associated expectation values. In \cite{CHtimelike} these expectation values were stated
without proof. One of the aims of this paper is to present a derivation for these results.

There are several ways to arrive at the action of generators on states of unitary irreps of $\SLC$.
The first approach is algebraic and matrix elements are inferred by repeated use of the
$\SLC$ commutation relations.

This is the method by which Gelfand and Naimark determined the matrix elements of $\SLC$ generators in the
SU(2) decomposition (see \cite{Naimark}). The same matrix elements were derived independently by 
Harish--Chandra \cite{HarishChandra}.
Sciarrino \& Toller \cite{SciarrinoToller} and Delbourgo et al.\ \cite{Delbourgo} investigated $\SLC$ matrix
elements from the perspective of the method of induced representations.
They deduced expressions for matrix elements of finite $\SLC$ transformations and for transition functions between SU(2) and
SU(1,1) states. By continuing from this point one could also determine infinitesimal expressions.
Another possibility is to start from the realization in terms of homogeneous functions of two complex variables (see \cite{Naimark,Ruhl,Carmeli})
and to evaluate the matrix elements of finite transformations by explicit integration.
For SU(2) this was done by Strom \cite{Strom1967vol33} and Duc \& Van Hieu \cite{DucVanHieu}.
Finally, one can take the differential approach---represent generators as differential operators and act on state functions
with them. This is the method by which Mukunda derived the matrix elements of $\SLC$ generators in the
SU(1,1) decomposition, for eigenstates of $J^3$ and integer spin \cite{Mukunda}.

This is also the strategy followed in the present paper.
While based on the same method, our results extend those of ref.\ \cite{Mukunda} in several ways.
In addition to the usual basis diagonalized by $J^3$, we compute the matrix elements
for a basis of $K^1$ eigenstates\footnote{As mentioned before, this is the type of states needed
to represent timelike quantum triangles.}.
Since $K^1$ generates a noncompact subgroup of SU(1,1), this basis is labelled by continuous eigenvalues.
Furthermore, we present both the treatment of the multiplicative and derivative part of the operator
and find certain corrections to Mukunda's result\footnote{In \cite{Mukunda} the proof is given
for the multiplicative term and the result for the total operator is only stated.}.
We also clarify the definition of the state functions by relating them directly to the $D$--functions
of SU(1,1) and SU(2).
By means of suitable parametrizations we are able to highlight the common structure present in differential
operators and state functions for different choices of basis. Thus, we can reduce the derivation to one
main equation and treat several cases at once.
With the inclusion of the canonical SU(2) basis we deal in total with five cases that are listed
in table \eq{tablecases}.

\renewcommand{\arraystretch}{2}
\begin{table}
\begin{tabular}{@{\qquad}l@{\qquad}|@{\qquad}l@{\qquad}|@{\qquad}l}
group & series & basis \\ \hline
SU(2)  & & $J^3$ \\
SU(1,1) & discrete & $J^3$ \\
SU(1,1) & continuous & $J^3$ \\
SU(1,1) & discrete & $K^1$ \\
SU(1,1) & continuous & $K^1$
\end{tabular}
\caption{\label{tablecases} Cases treated in this paper, listed according to group, series and diagonal operator. 
In the last line, eigenvalues of $K^1$ occur with multiplicity 2.}
\end{table}

The article is organized as follows. In section \ref{representationtheory} we briefly review
basic facts about representations of $\SLC$, SU(2) and SU(1,1)
that are needed to understand the rest of the paper.
In section \ref{basisstates} we give explicit definitions of the basis states
used to define matrix elements of the $\SLC$ representation.
The main result of the paper is stated in \sec{matrixelementsofSL2Cgenerators}:
the matrix elements of $\SLC$ generators in the SU(1,1) decomposition---
in a discrete and continuous basis and for both discrete and continuous series.
The derivation of the matrix elements is explained, in some detail,
in \sec{derivation}.
This section also refers to the appendix, where we provide
further details on the parametrization of the groups (\sec{parametrization}),
the definition of the Bargmann functions (\sec{representationfunctionsofSU(1,1)})
and on the derivation of the main equation of the paper (\sec{derivationofmainequation}).
We conclude with a brief summary and discussion (\sec{summary}).

\section{Representation theory of $\SLC$, SU(2) and SU(1,1)}
\label{representationtheory}

\renewcommand{\arraystretch}{1.3}
In the defining representation, $\mathrm{SL(2,\bC)}$ has the generators $J^i = \sigma^i/2$, $K^i = \irm\sigma^i/2$, $i = 1,2,3$,
with commutation relations
\be
[J^i,J^j] = \irm \epsilon^{ijk} J^k\,,\qquad [J^i,K^j] = \irm \epsilon^{ijk} K^k\,,\qquad [K^i,K^j] = -\irm \epsilon^{ijk} K^k\,.
\label{commutationrelations}
\ee
The subgroups SU(2) and SU(1,1) are generated by $J^1, J^2, J^3$ and $J^3, K^1, K^2$ respectively.
$\v{J}$ and $\v{K}$ transform as vectors under SU(2). For SU(1,1) an analogous role is played by the vectors
$\v{F} \equiv (J^3,K^1,K^2)$ and $\v{G} \equiv (K^3,-J^1,-J^2)$, which transform as Minkowski vectors under SU(1,1) \cite{Mukunda}.

Unitary irreps of $\mathrm{SL(2,\bC)}$ are labelled by pairs of numbers $(\rho,n)$, $\rho\in\bR$, $n\in\bZ_+$,
which are related to the two Casimirs $C_1$ and $C_2$:
\bea
C_1 &=& 2\left(\v{J}^2 - \v{K}^2\right) = \frac{1}{2} (n^2 - \rho^2 - 4)\,, \\
C_2 &=& -4 \v{J}\cdot \v{K} = n\rho\,.
\eea
The representation space $\clH_{(\rho,n)}$ consists of functions $F: \bC^2\backslash\{0\} \to \bC$ with the homogeneity property
\be
F(\alpha z_1,\alpha z_2) = \alpha^{\irm \rho/2 + n/2 - 1} {\alpha^*}^{\irm \rho/2 - n/2 - 1} F(z_1,z_2) \quad \forall\;\alpha\in\bC\backslash\{0\}\,.
\label{homogeneity}
\ee
The representation acts on these functions by
\be
D^{(\rho,n)}(g) F(z_1,z_2) = F(a z_1 + c z_2, b z_2 + d z_2)\,,\qquad g\in\twomatrix{a}{b}{c}{d}\in \mathrm{SL(2,\bC)}\,.
\label{SL2Crepresentation}
\ee
The inner product of $\clH_{(\rho,n)}$ is constructed from the $\SLC$--invariant 2--form
\be
\omega = \frac{\irm}{2} (z_2 \d z_1 - z_1 \d z_2)\wedge (\zb_2 \d\zb_1 - \zb_1 \d\zb_2)\,.
\ee
For homogeneous functions $F_1, F_2$ of the type \eq{homogeneity}, the 2--form $F_1^* F_2 \omega$ is invariant under $(z_1,z_2) \to (\lambda z_1, \lambda z_2)$,
$\lambda\in\bC\backslash\{0\}$. Thus, $F_1^* F_2 \omega$ projects to a 2--form under $\pi: \bC^2\backslash\{0\}\to \bC P^1$
and one can specify the inner product by
\be
\b F_1 | F_2 \ket \equiv \int\limits_{\bC P^1} \pi(F_1^* F_2 \omega)\,.
\label{innerproductC^2}
\ee
Since $\omega$ is $\SLC$--invariant, the representation is unitary w.r.t.\ this inner product.
Equation \eq{innerproductC^2} can be equivalently expressed in terms of sections of the bundle $\bC^2\backslash\{0\}\to \bC P^1$. In particular, when choosing
the section $z \mapsto (z,1)$, one obtains
the integral
\be
\b F_1 | F_2 \ket = \int \d x\,\d y\; F_1^*(z,1) F_2(z,1)\,,\qquad z = x + \irm y\,.
\ee

The unitary irreps of SU(2) and SU(1,1) can be built from eigenstates $|j\, m\ket$ of $J^3$:
\be
J^3\, |j\, m\ket = m |j\, m\ket\,,\qquad \b j\, m | j\, m'\ket = \delta_{mm'}\,.
\label{J3eigenstate}
\ee
In the case of SU(2), the irreps are labelled by the Casimir $\v{J}^2$:
\be
\v{J}^2\, |j\, m\ket = j(j+1) |j\, m\ket\,,\qquad\mbox{where $j = k/2$, $k\in\bN_0$.}
\ee
The representation space $\clD_j$ of spin $j$ consists of states with $m = -j,\ldots, j$.
The raising and lowering operators are given by
\be
J^\pm = J^1 \pm \irm J^2\,,\qquad J^\pm\,|j\, m\ket = \sqrt{(j \pm m + 1)(j \mp m)}\,|j\, m\pm 1\ket\,.
\ee

Unitary irreps of SU(1,1) have the Casimir $Q = \v{F}^2 = (J^3)^2 - (K^1)^2 - (K^2)^2$,
\be
Q\, |j\, m\ket = j(j+1) |j\, m\ket\,,
\ee
and split into two classes, the discrete series and the continuous series.
For the discrete series, the spin $j$ assumes negative values $j = -k/2$, $k \in \bN$.
Irreps of the \textit{positive} (\textit{negative}) discrete series are denoted by $\clD^\pm_j$ and consist of states $|j\, m\ket$
with eigenvalues $m = -j,\; -j+1,\; -j+2,\; \ldots$ and $m = j,\; j-1,\; j-2,\; \ldots$ respectively.
In the case of the continuous series, the spin $j$ is complex and the Casimir has a continuous spectrum:
\be
Q\, |j\, m\ket = j(j+1) |j\, m\ket\,, \qquad\mbox{where $j = -\frac{1}{2} + \irm s$,\quad $0 < s < \infty$\,.}
\ee
Irreps of this series are denoted by $\clC^\epsilon_s$. The allowed values for $m$ are either
\be
m = 0,\, \pm 1,\, \pm 2,\, \ldots\qquad\mbox{or}\qquad m = \pm\frac{1}{2},\, \pm\frac{3}{2},\, \ldots\,,
\ee
and the label $\epsilon = 0, \frac{1}{2}$ designates these two possibilities.
In both the discrete and continuous series raising and lowering is achieved by the operators
\be
F^\pm = F^2 \mp \irm F^1\,,\qquad F^\pm |j\, m\ket = \sqrt{(m \pm j \pm 1)(m \mp j)}\, |j\, m\pm 1\ket\,.
\label{raisingloweringSU(1,1)}
\ee

As an alternative to the $|j\, m\ket$ basis one can use eigenstates of $K^1$ (see \cite{LindbladNagel} for details and \cite{MukundaO21inO11basis} for early work
in this direction):
\be
K^1\, |j\,\lambda\,\sigma\ket = \lambda |j\,\lambda\,\sigma\ket\,,\qquad \b j\,\lambda'\,\sigma' |j\,\lambda\,\sigma\ket = \delta(\lambda' - \lambda) \delta_{\sigma'\sigma}\,.
\label{K1eigenstate}
\ee
Since $K^1$ generates a noncompact subgroup, these eigenstates are not normalizable.
In the continuous series there occurs a two--fold degeneracy of the spectrum which is labelled
by the additional index $\sigma = 0,1$.
In analogy to \eq{raisingloweringSU(1,1)} one may define ``raising'' and ``lowering'' operators 
\be
\clF^\pm = F^0 \mp F^2\,,\qquad \clF^\pm |j\,\lambda\,\sigma\ket = \irm (\pm j \pm 1 - \irm\lambda)\, |j\,(\lambda\pm\irm)\,(\sigma + 1\!\!\!\!\mod 2)\ket\,.
\label{"raisinglowering"SU(1,1)}
\ee
The shift $\lambda \pm \irm$ to complex eigenvalues follows from the commutation relations, but how is this consistent with $K^1$ being a self--adjoint operator? The answer is related to the fact that eigenvectors of $K^1$ are described as elements of a dual space $\clD'$ in a Gelfand triple $\clD \subset \clH \subset \clD'$. The operator $K^1$ is self--adjoint in the Hilbert space $\clH$ and has the generalized eigenvectors $|j\,\lambda\,\sigma\ket\in \clD'$ with real eigenvalues $\lambda$. However, when extended to $\clD'$, the operator $K^1$ has eigenvectors for all complex $\lambda$ and the state $\clF^\pm |j\,\lambda\,\sigma\ket\in \clD'$ is an example of such an eigenvector.

\section{Basis states for unitary irreps of $\SLC$}
\label{basisstates}

\subsection{SU(2) and SU(1,1) decomposition}

Clearly, every unitary irrep of $\mathrm{SL(2,\bC)}$ defines a representation of its subgroups SU(2) and SU(1,1).
However, these representations are reducible. As a result, the Hilbert space $\clH_{(\rho,n)}$ splits into
a direct sum of irreps of SU(2), or a direct sum of irreps of SU(1,1) \cite{Ruhl}.
The SU(2) decomposition is given by the following isomorphism and completeness relation:
\be
\clH_{(\rho,n)} \simeq \bigoplus\limits_{j = n/2}^{\infty} \clD_j\,,\qquad
\mathbbm{1}_{(\rho,n)} = \sum\limits_{j = n/2}^\infty \sum_{m=-j}^j \left|\Psi_{j\, m}\right\ket \left\b\Psi_{j\, m}\right|\,.
\label{decompositionSU(2)}
\ee
The states $|\Psi_{j\, m}\ket$ form the so--called canonical basis of $\clH_{(\rho,n)}$.
For fixed spin $j$ and $m = -j,\ldots, j$, they span a subspace of $\clH_{(\rho,n)}$ that is isomorphic
to $\clD_j$. The SU(1,1) reduction can be written as
\be
\clH_{(\rho,n)} \quad\simeq\quad
\left(\bigoplus\limits_{j < -1/2}^{-n/2} \clD^+_j \oplus\!\!\!\!\! \int\limits_0^{\;\;\;\;\;\infty\; \oplus} \!\!\!\!\!\d s\; \clC^\epsilon_s\right)
\oplus
\left(\bigoplus\limits_{j < -1/2}^{-n/2} \clD^-_j \oplus\!\!\!\!\! \int\limits_0^{\;\;\;\;\;\infty\; \oplus} \!\!\!\!\!\d s\; \clC^\epsilon_s\right)\,.
\label{decompositionSU(1,1)}
\ee
The precise meaning of this statement is encoded in the completeness relation
\be
\mathbbm{1}_{(\rho,n)} =
\sum_{\tau = \pm 1} \left\{
\sum\limits_{j < -1/2}^{-n/2} \sum_{\tau m=-j}^{\infty} \left|\Psi^\tau_{j\, m}\right\ket \left\b\Psi^\tau_{j\, m}\right|
+
\int\limits_0^\infty \d s\;\mu_\epsilon(s) \sum\limits_{\pm m = \epsilon}^\infty \left|\Psi^\tau_{j\, m}\right\ket \left\b\Psi^\tau_{j\, m}\right|
\right\}\,.
\label{completenessrelationSU(1,1)}
\ee
Here, the states $|\Psi^\tau_{j\, m}\ket$, $-\frac{n}{2} \le j < -\frac{1}{2}$, and $|\Psi^\tau_{j\, m}\ket$, $j = -\frac{1}{2} + \irm s$, correspond to states $|j\, m\ket$
of the discrete and continuous series respectively.
The sum over $j$ extends over values such that $j - n/2$ is integral. Moreover, $\epsilon$ has a value such that $\epsilon - n/2$ is an integer.
The measure factor $\mu_\epsilon(s)$ depends on the specific choice of normalization for the states $|\Psi^\tau_{j\, m}\ket$ and
will be given below.

\renewcommand{\arraystretch}{1.5}
The decompositions \eq{decompositionSU(2)} and \eq{decompositionSU(1,1)} can be derived from the
homogeneity property \eq{homogeneity} and the Plancherel decomposition of SU(2) and SU(1,1) respectively.
Due to \eq{homogeneity} every function $F$ in $\clH_{(\rho,n)}$ can be equivalently described by a function $f$ of SU(2) via
\be
F(z_1,z_2) = \sqrt{\pi} \left(|z_1|^2 + |z_2|^2\right)^{\irm\rho/2 - 1} f(u(z_1,z_2))\,,\qquad
u(z_1,z_2) = \frac{1}{\sqrt{|z_1|^2 + |z_2|^2}}\twomatrix{z_1}{z_2}{-\zb_2}{\zb_1}\,.
\label{FasSU(2)function}
\ee
Thus, the Hilbert space $\clH_{(\rho,n)}$ is isomorphic to a subspace of $L^2(SU(2))$. Under this isomorphism the inner product \eq{innerproductC^2}
turns into
\be
\b f_1 | f_2\ket = \int \d u\; f^*_1(u) f_2(u)\,,
\ee
where $\d u$ denotes the normalized Haar measure on SU(2).
Alternatively, the functions $F$ can be characterized by pairs $(f^+,f^-)$ of functions $f^\tau: \mathrm{SU(1,1)}\to \bC$, $\tau = \pm 1$,
via
\be
F(z_1,z_2) = \sqrt{\pi} \left(\tau |z_1|^2 - \tau |z_2|^2\right)^{\irm\rho/2 - 1} f^\tau(v^\tau(z_1,z_2))\,,\qquad
\tau =
\left\{
\begin{array}{rl}
1\,, & |z_1| > |z_2|\,, \\
-1\,, & |z_1| < |z_2|\,,
\end{array}
\right.
\label{FasSU(1,1)function}
\ee
where we choose
\be
v^\tau(z_1,z_2) =
\left\{
\begin{array}{rl}
\frac{1}{\sqrt{|z_1|^2 - |z_2|^2}}
\twomatrix{z_1}{z_2}{\zb_2}{\zb_1}\,,
& \tau = 1\,, \\
& \\
\frac{1}{\sqrt{|z_2|^2 - |z_1|^2}}
\twomatrix{\zb_2}{\zb_1}{z_1}{z_2}\,,
& \tau = -1\,.
\end{array}
\right.
\ee
As a result, $\clH_{(\rho,n)}$ is isomorphic to a subspace of $L^2(SU(1,1))\oplus L^2(SU(1,1))$ with the inner product
\be
\left\b \left(f^+_1,f^-_1\right) \left| \left(f^+_2,f^-_2\right)\right\ket\right. = \sum_{\tau = \pm 1} \int \d v\; \left(f^\tau_1(v)\right)^* f^\tau_2(v)\,.
\ee
The measure $\d v$ is specified in appendix \ref{parametrization}.

\renewcommand{\arraystretch}{2.3}
Functions of SU(2) or SU(1,1) can be expanded in matrix elements
\be
D^j_{x' x}(g) \equiv \b j\,x'| D^j(g) |j\,x\ket\,,
\ee
where $x'$ and $x$ label appropriate basis states.
When applied to the functions $f$ on SU(2), this leads to the decomposition \eq{decompositionSU(2)} with states given by the basis functions
\be
\Psi_{j\,m}(u) = \sqrt{2j+1}\, D^j_{n/2\, m}(u)\,.
\label{stateSU(2)}
\ee
Similarly, the SU(1,1) decomposition results in \eq{decompositionSU(1,1)} with states represented by functions
\be
\Psi^\tau_{j\,x}(v)
=
\left\{
\begin{array}{ll}
\sqrt{2j+1} \left(D^j_{n/2\, x}(v),0\right)\,, & \tau = 1\,, \\
\sqrt{2j+1} \left(0,D^j_{-n/2\, x}(v)\right)\,, & \tau = -1\,,
\end{array}
\right.
\qquad j\neq -1\,.
\label{stateSU(1,1)}
\ee
The label $x$ is $m$ if we use a basis of $J^3$ eigenstates. When employing $K^1$ eigenstates,
one has $x = \lambda$ for the discrete series and $x = \lambda\,\sigma$ for the continuous series.
The label $x'$ is set to $\pm n/2$ corresponding to the state $|j\,\pm \! n/2\ket$. 
The choice of normalization \eq{stateSU(1,1)} determines the measure factor
$\mu_\epsilon(s)$ in \eq{decompositionSU(1,1)} to be \renewcommand{\arraystretch}{1.8}
\be
\mu_\epsilon(s) = \left\{
\begin{array}{ll}
\ds -\irm\tanh(\pi s)\,, & \epsilon = 0\,, \\
\ds -\irm\coth(\pi s)\,, & \epsilon = 1/2\,.
\end{array}
\right.
\ee
The irrep of spin $j = -1/2$ represents a special case that does not appear in the Plancherel decomposition.
However, since it will come up in calculations below, we define the associated state
\be
\Psi^\tau_{j\,x}(v)
\equiv
\left\{
\begin{array}{ll}
\left(D^j_{n/2\, x}(v),0\right)\,, & \tau = 1\,, \\
\left(0,D^j_{-n/2\, x}(v)\right)\,, & \tau = -1\,.
\end{array}
\right.
\label{stateSU(1,1)j=-1/2}
\ee

\subsection{Explicit expressions for basis functions}
\label{explicitexpressions}

In order to derive the action of generators on states in \sec{matrixelementsofSL2Cgenerators}
we need explicit expressions for the state functions \eq{stateSU(2)} and \eq{stateSU(1,1)}.
Altogether we will encounter five different cases, depending on the group, the choice of basis states
and the series (see table \ref{tablecases}).
When dealing with the associated $D$--functions we will exploit the fact
that they all share a similar structure. In each case, the $D$--function can be built from
the expression
\be
F^j_{m'm}(z) = (1 - z)^{(m'+m)/2} z^{(m'-m)/2}\, {}_2F_1(-j + m', j + m' + 1, m' - m + 1; z)
\label{Fjm'm}
\ee
where ${}_2F_1$ denotes Gauss' hypergeometric function. The full $D$--functions are obtained
by including normalization factors, phases and a suitable parametrization of $z$.

Let us start with the group SU(2). When using the parametrization \eq{parametrizationSU(2)}
the $D$--function of SU(2) reduces to the Wigner $d$--function via
\be
D^j_{m'm}(u) = \e^{\irm m' \psi}\,d^j_{m'm}(\theta)\,\e^{\irm m\varphi}\,.
\ee
In the case $m' \ge m$, $m' + m \ge 0$, the $d$--function has the explicit form
\be
d^j_{m'm}(\theta) = \frac{1}{(m' - m)!}\, N^j_{m'm} F^j_{m'm}(z(\theta))
\label{dfunctionSU(2)}
\ee
where
\be
z(\theta) \equiv \frac{1}{2}(1 - \cos\theta)
\ee
and $F^j_{m'm}(z)$ is the function defined in \eq{Fjm'm} \cite{AndrewsGunson}. \setlength{\jot}{0.3cm}
The normalization factor $N^j_{m'm}$ can be written in several ways:
\bea
N^j_{m'm}
&=&
\left[\prod_{l = 0}^{m' - m - 1} (j + m' - l)(j - m - l)\right]^\half \\
&=&
\left[\frac{(j + m')!(j - m)!}{(j + m)!(j - m')!}\right]^\half
=
\left[\frac{\Gamma(j + m' + 1)\Gamma(j - m + 1)}{\Gamma(j + m + 1)\Gamma(j - m' + 1)}\right]^\half
\eea
The expressions for other values of $m'$ and $m$ follow from table \ref{tabledfunction}.

\begin{table}
$
\begin{array}{l@{\quad}l@{\quad}|@{\hspace{1.3cm}}r@{\quad}|@{\quad}r}
& & m' \ge m & m' < m \\ \hline
\mbox{SU(2)} & m' + m \ge 0 & d^j_{m',m}(\theta) & (-1)^{m'-m} d^j_{m,m'}(\theta) \\
& m' + m \le 0 & d^j_{-m,-m'}(\theta) & (-1)^{m'-m} d^j_{-m',-m}(\theta) \\ \hline
\mbox{SU(1,1) discrete} & m', m \ge -j & b^j_{m',m}(t) & b^j_{m,m'}(t) \\
& m', m \le j & b^j_{-m,-m'}(t) & b^j_{-m',-m}(t) \\ \hline
\mbox{SU(1,1) continuous} & & b^j_{m',m}(t) & b^j_{m,m'}(t)
\end{array}
$
\caption{\label{tabledfunction} The Wigner functions $d^j_{m',m}(\theta)$ obey symmetries
that allow one to infer its values for general $m'$, $m$ from those for $m' \ge m$, $m' + m \ge 0$.
For example, for $m' < m$ and $m' + m \le 0$, $d^j_{m'm} = (-1)^{m'-m} d^j_{-m',-m}$. Similar rules
apply to the Bargmann function $b^j_{m',m}(t)$.}
\end{table}

Next consider the case of SU(1,1) with a basis of $J^3$ eigenstates.
Under the parametrization \eq{parametrizationSU(1,1)J^3} we have,
for both the discrete and continuous series, that
\be
D^j_{m'm}(v) = \e^{\irm m' \psi}\,b^j_{m'm}(t)\,\e^{\irm m\varphi}
\label{DfunctionSU(1,1)J^3}
\ee
where $b^j_{m'm}(t)$ is an SU(1,1) analog of the Wigner $d$--function.
The explicit form of the $b$--functions was determined by Bargmann \cite{Bargmann}.
For $m' \ge m$, $m' + m \ge 0$, these can be written as
\be
b^j_{m'm}(t)
= \sqrt{(-1)^{m - m'}}\, d^j_{m'm}(\irm t)
= \frac{1}{(m' - m)!}\, \Nt^j_{m'm} F^j_{m'm}(z(\irm t))
\label{bfunctionSU(1,1)J^3}
\ee
where the normalization factor is given by
\be
\Nt^j_{m'm}
= \left[\prod_{l = 0}^{m' - m - 1} (j + m' - l)(m - j + l)\right]^\half \\
= \left[\frac{\Gamma(m' + j + 1)\Gamma(m' - j)}{\Gamma(m + j + 1)\Gamma(m - j)}\right]^\half\,.
\label{normalizationSU(1,1)J^3}
\ee
The other cases are obtained from table \ref{tabledfunction}.
In appendix \ref{representationfunctionsofSU(1,1)} it is shown that this definition
is indeed equivalent to the one provided by Bargmann.

Finally, we come to SU(1,1) and a basis of $K^1$ eigenstates.
According to eq.\ \eq{stateSU(1,1)} we need expressions for $D$--functions in ``mixed'' bases, 
where the left state belongs to the discrete $J^3$ basis and
the right state is from the continuous basis diagonal in $K^1$.
This case was worked out by Lindblad \cite{Lindblad}, using previous results by Lindblad and Nagel \cite{LindbladNagel}.
For the discrete series and the parametrization \eq{parametrizationSU(1,1)K^1},
\be
D^j_{m\lambda}(v) = \e^{\irm m \varphi}\,d^j_{m\lambda}(t)\,\e^{\irm \lambda u}
\ee
and for $m\ge -j$ \renewcommand{\arraystretch}{2.3}
\be
d^j_{m\lambda}(t) = N^j_m F^j_{-m,-\irm\lambda}(z(-t))\,.
\label{dfunctionSU(1,1)K^1discrete}
\ee  
Here, $z$ is parametrized by
\be
z(t) \equiv \frac{1}{2}(1 - \irm\sinh t)
\ee
and the normalization factor is defined by
\be
N^j_m \equiv
\frac{\sqrt{2}}{\pi} 2^{-j-2} S^j_m R^j_{m\lambda}\,,\qquad S^j_m \equiv \frac{\left[\Gamma(m - j)\Gamma(m + j + 1)\right]^{\half}}{\Gamma(m + j + 1)}\,,
\label{normalizationK^1discrete}
\ee
and
\be
R^j_{m\lambda} \equiv
\frac{\Gamma(j + 1 + \irm\lambda)\,\Gamma\left(\ds\frac{-j - \irm\lambda}{2}\right)\Gamma\left(\ds\frac{-j + 1 + \irm\lambda}{2}\right)}{\Gamma(m - j)\Gamma(-m + 1 + \irm\lambda)}\,.
\ee
The $d$--function for $m \le j$ results from
\be
d^j_{m\lambda}(t) = d^j_{-m\lambda}(-t)\,.
\label{dfunctionK^1discrete_mnegative}
\ee

For the continuous series, one has
\be
D^j_{m\lambda\sigma}(v) = \e^{\irm m \varphi}\,d^j_{m\lambda\sigma}(t)\,\e^{\irm \lambda u}
\ee
with the $d$--function
\be
d^j_{m\lambda\sigma}(t) = S^j_m\left[T^j_{m\lambda\sigma} F^j_{m,-\irm\lambda}(z(t)) - (-1)^\sigma T^j_{-m\lambda\sigma} F^j_{-m,-\irm\lambda}(z(-t))\right]\,.
\label{dfunctionSU(1,1)K^1continuous}
\ee 
The factor $S^j_m$ is specified as in \eq{normalizationK^1discrete} and
\be
T^j_{m\lambda\sigma} = \frac{2^{j - 1} \Gamma(-j + \irm\lambda)}{\irm^\sigma \sin\left[\frac{\pi}{2}(-j + \sigma - \irm\lambda)\right] \Gamma(-m - j)\Gamma(m + 1 + \irm\lambda)}\,.
\ee
The above formulas are identical to Lindblad's
except for sign switches due to differing conventions ($t\to -t$ and $\lambda \to -\lambda$).


\section{Matrix elements of $\SLC$ generators}
\label{matrixelementsofSL2Cgenerators}

In this section state we state our results---the matrix elements of $\SLC$ generators
in discrete and continuous bases of SU(1,1).
To save space we write down only the formula for one of the generators outside SU(1,1).
In the case of the $J^3$ basis, we choose the generator $K^3$ and in the case
of the $K^1$ basis we select $J^1$.
Given the matrix elements of $K^3$ (or $J^1$), the entire set of matrix elements can then
be readily computed from the commutation relations \eq{commutationrelations} and
the known action of generators of SU(1,1) (see eqns.\ \eq{J3eigenstate}, \eq{raisingloweringSU(1,1)}
\eq{K1eigenstate} and \eq{"raisinglowering"SU(1,1)}).
For completeness we also include the result for the subgroup SU(2).
The derivation of the different cases is presented in section \ref{derivation}.
Each of the subsequent formulas has been checked numerically for a number of parameter values.

\setlength{\jot}{0.3cm}
Let us define coefficients
\be
A_j = \frac{\rho\,n}{4 j(j+1)}\,,\qquad
C_j = \frac{\sqrt{n^2/4 - j^2} \sqrt{m^2 - j^2}}{j \sqrt{2j - 1}\sqrt{2j+1}}\,,
\ee
For the canonical SU(2) decomposition the action of $K^3$ is well--known
\cite{Naimark,Carmeli} and we state it here for the explicit choice of states given in eq.\
\eq{stateSU(2)}. For $j\neq 0, \textstyle\frac{1}{2}$, one has
\be
K^3\,\big|\Psi_{j\,m}\big\ket =
\big[\rho/2 + \irm(j+1)\big] C_{j+1}\, \big|\Psi_{j+1\,m}\big\ket
- m A_j\, \big|\Psi_{j\,m}\big\ket
+ (\rho/2 - \irm j) C_j\, \big|\Psi_{j-1\,m}\big\ket
\label{K^3SU(2)}
\ee 
In the case of $j=1/2$, the third term on the right--hand side is absent and for $j=0$ the second and third term are absent. 
This formula differs slightly from the one in \cite{Naimark},
since the algebraic derivation assumes a suitable choice of phase in the states, dependent
on $\rho$ and $j$, so that $\rho/2 + \irm(j+1)$ becomes $\sqrt{(\rho/2)^2 + (j+1)^2}$ etc.\footnote{See
the remarks on choice of phase in Tung's textbook \cite{Tung}, sec.\ 10.3.3 and appendix VII.}

Next consider the SU(1,1) decomposition w.r.t.\ $J^3$ eigenstates. For the discrete series $j \le -2$
and the continuous series,\footnote{For integer spin, this case was previously derived by Mukunda \cite{Mukunda}.
We find, however, minor discrepancies with our result, which can be traced back to parts of the
proof that were not presented in \cite{Mukunda} (see appendix \ref{derivationofmainequation}).
For example, $(\rho - \irm (k - 1))$ in eq.\ (3.19) \cite{Mukunda} should be replaced by $(\rho + \irm (k - 1))$.}
\be
K^3\,\big|\Psi^\tau_{j\,m}\big\ket =
\tau \big[\rho/2 + \irm(j+1)\big] C_{j+1}\,\big|\Psi^\tau_{j+1\,m}\big\ket
- m A_j\, \big|\Psi^\tau_{j\,m}\big\ket
+ \tau (\rho/2 - \irm j) C_j\, \big|\Psi^\tau_{j-1\,m}\big\ket
\label{K^3J^3states}
\ee 
The equation for $j = -3/2$ is special, since then $j+1 = -1/2$, leading to the state \eq{stateSU(1,1)j=-1/2} outside the
Plancherel decomposition. In this case, the denominator $(j+1) \sqrt{2j + 1}\sqrt{2j + 3}$ in $C_{j+1}$ has to be replaced
by $(j+1)\sqrt{2j+1}$.
Similar ``boundary'' effects occur for $j = -1$, where the $K^3$ action can be cast in the form
\bea
K^3\,\big|\Psi^\tau_{-1\,m}\big\ket &=&
\frac{1}{2}\,\rho m n\,\pa_j\big|\Psi^\tau_{j\,m}\big\ket\big|_{j = -1} \nonumber \\
&& {}+ \frac{1}{2}\,(\rho/2 - \irm) m n \,\big|\Psi^\tau_{-1\,m}\big\ket \nonumber \\
&& {}+ \frac{1}{\sqrt{3}}\tau\, (\rho/2 + \irm) \sqrt{n^2/4 - 1} \sqrt{m^2 - 1} \,\big|\Psi^\tau_{-2\,m}\big\ket\,.
\label{j=-1J^3basis}
\eea 
The $j+1 = -1/2$ term for $j = 3/2$ and the first term in \eq{j=-1J^3basis}
are related to states of the continuous series through analytic continuation (see the discussion in \cite{Mukunda}).

\setlength{\jot}{0.3cm}
Coming to the SU(1,1) decomposition w.r.t.\ $K^1$ eigenstates, it is convenient to specify coefficients
\be
\Ct_j = \frac{\sqrt{n^2/4 - j^2}}{j \sqrt{2j - 1} \sqrt{2j+1}}\,,\qquad
B_j = \Gamma\left(\frac{-j-\irm\lambda}{2}\right) \Gamma\left(\frac{-j+1+\irm\lambda}{2}\right)\,.
\ee
In the case of the discrete series and $j \le -2$, the action of $J^1$ gives
\bea
J^1\,\big|\Psi^\tau_{j\,\lambda}\big\ket &=&
2\irm \big[\rho/2 + \irm(j+1)\big] \frac{B_j}{B_{j+1}}\, \Ct_{j+1}\, \big|\Psi_{j+1\,\lambda}\big\ket \nonumber \\
&& {}+ \lambda A_j\, \big|\Psi_{j\,\lambda}\big\ket \nonumber \\
&& {}- \frac{\irm}{2}\, (\rho/2 - \irm j) (j^2 + \lambda^2) \frac{B_j}{B_{j-1}}\, \Ct_j\, \big|\Psi_{j-1\,\lambda}\big\ket
\label{J^1discrete}
\eea 
As before, the action on the state $j = -3/2$ results in a state $j+1 = -1/2$, with the factor $(j+1) \sqrt{2j + 1}\sqrt{2j + 3}$ in $\Ct_{j+1}$
substituted by $(j+1)\sqrt{2j+1}$. For brevity, we do not spell out the case $j = -1$, which produces a formula similar to eq.\ \eq{j=-1J^3basis}.
On continuous series states $J^1$ yields, for $n\neq 0$, \setlength{\jot}{0.5cm}
\bea
J^1\,\big|\Psi^\tau_{j\,\lambda\,\sigma}\big\ket &=&
-\frac{1}{2} \big[\rho/2 + \irm(j+1)\big] \left[(j+1)^2 + \lambda^2\right] \Ct_{j+1}\, \big|\Psi^\tau_{j+1\,\lambda\,\sigma'}\big\ket \nonumber \\
&& {}+ \lambda A_j\, \big|\Psi^\tau_{j\,\lambda\,\sigma}\big\ket \nonumber \\
&& {}- 2\, (\rho/2 - \irm j) \Ct_j\, \big|\Psi^\tau_{j-1\,\lambda\,\sigma'}\big\ket\,,
\label{J^1continuous}
\eea
where $\sigma' = \sigma + 1 \bmod 2$. When $n = 0$, the right--hand side comes with an additional factor $\tau$ in front of the $j+1$ term.

In all of the previous equations, $\Delta j = \pm 1, 0$,
in accordance with the Wigner--Eckart theorem and the fact that $K^3$ and $J^1$ are components of vector operators.
Since the vectors $\v{F} = (J^3,K^1,K^2)$ and $\v{G} = (K^3,-J^1,-J^2)$ transform as Minkowski vectors
under SU(1,1), the associated Clebsch--Gordan coefficients correspond to the coupling of unitary SU(1,1) irreps
with the \textit{non--unitary} SU(1,1) irrep of spin 1 (see \cite{Ui}).

\section{Derivation of matrix elements}
\label{derivation}

In this section, we outline how the matrix elements of the previous section were obtained.
The first step consists in expressing the generators of $\SLC$ as differential
operators of the relevant subgroup (SU(1,1) or SU(2)). In doing so we employ parametrizations that are adapted
to the choice of basis states (either $J^3$ or $K^1$ eigenstates).
Once the differential operators are determined, we apply them to the state functions defined in
\sec{explicitexpressions}. More precisely, we act with $K^3$ on the basis of $J^3$
eigenstates and with $J^1$ on the basis diagonalized by $K^1$. The resulting states
are decomposed with respect to the original basis, thus giving us the matrix elements
of $K^3$ and $J^1$ respectively.

\subsection{Generators as differential operators}
\label{differentialoperators}

In order to derive the differential operators associated to $\SLC$ generators,
we start from the definition of the representation \eq{SL2Crepresentation} and combine it
with the relation \eq{FasSU(1,1)function} between functions $F$ of $\bC^2\backslash\{0\}$ and functions $f^\tau$ of SU(1,1)
to get the finite transformation of $f^\tau$:
\renewcommand{\arraystretch}{1.5}
\be
D^{(\rho,n)}(g) f^\tau(v^\tau) = \left(\tau |a v_1 + c v_2|^2 - \tau |b v_1 + d v_2|^2\right)^{\irm \rho/2 - 1} f^\tau(v^\tau\cdot g)
\label{finitetransformation}
\ee
where
\be
v^\tau \equiv
\left\{
\begin{array}{rl}
\twomatrix{v_1}{v_2}{\vbar_2}{\vbar_1}\,,
& \tau = 1\,, \\
& \\
\twomatrix{\vbar_2}{\vbar_1}{v_1}{v_2}\,,
& \tau = -1\,,
\end{array}
\right.
\qquad \mbox{and} \qquad \renewcommand{\arraystretch}{4.5}
\raisebox{0.4cm}{
$
\begin{array}{lcl}
\ds v_1\cdot g &=& \ds\frac{a v_1 + c v_2}{\left(\tau |a v_1 + c v_2|^2 - \tau |b v_1 + d v_2|^2\right)^{\frac{1}{2}}}\,, \\
\ds v_2\cdot g &=& \ds\frac{b v_1 + d v_2}{\left(\tau |a v_1 + c v_2|^2 - \tau |b v_1 + d v_2|^2\right)^{\frac{1}{2}}}\,.
\end{array}
$
}
\ee
From this we obtain the corresponding infinitesimal operators via
\be
J^i f^\tau = -\irm \pa_\epsilon \left.\left[D^{(\rho,n)}(a_i(\epsilon)) f^\tau(v^\tau)\right]\right|_{\epsilon = 0}\,,\qquad
K^i f^\tau = -\irm \pa_\epsilon \left.\left[D^{(\rho,n)}(b_i(\epsilon)) f^\tau(v^\tau)\right]\right|_{\epsilon = 0}\,.
\ee
Here, $a_i(\epsilon)$ and $b_i(\epsilon)$ stand for the group elements generated by $J^i$ and $K^i$, as defined in appendix \ref{parametrization}.
Analogous formulas hold for the SU(2) case.

When the generator resides in su(1,1), the transformation \eq{finitetransformation} is the natural action of SU(1,1) on functions of SU(1,1) and
the associated differential operator can be determined by standard methods (see e.g.\ \cite{Byrd}).
When the generator lies outside of su(1,1) (like $K^3$), the prefactor in \eq{finitetransformation} leads to a multiplicative term
and the infinitesimal transformation of the function's argument turns into a linear combination of SU(1,1) generators.
By substituting the differential expressions for these, one arrives at the differential operator for the $\SLC$ generator.

Let us list the results for the different cases and parametrizations.
In the case of the subgroup SU(2), where we use a $J^3$ basis and the parametrization \eq{parametrizationSU(2)}, we have
\bea
J^3 &=& -\irm\pa_\varphi\,, \\
J^\pm &=& \irm\,\e^{\pm\irm\varphi}\left(\cot\theta\,\pa_\varphi \mp \irm\pa_\theta - \frac{1}{\sin\theta}\,\pa_\psi\right)\,, \\
K^3 &=& -(\rho/2 + \irm) \cos\theta - \irm\sin\theta\,\pa_\theta\,.
\eea 
For SU(1,1) with a $J^3$ basis and parametrization \eq{parametrizationSU(1,1)J^3},
\bea
J^3 &=& -\irm\pa_\varphi\, \mathbbm{1}\,, \\
F^\pm &=& \pm\, \e^{\pm\irm\varphi}\left(\irm\coth t\,\pa_\varphi \mp \irm\pa_t - \frac{1}{\sinh t}\,\pa_\psi\right) \mathbbm{1}\,, \\
K^3 &=& \big[-(\rho/2 + \irm) \cosh t - \irm\sinh t\,\pa_t\big]\,\sigma_3\,.
\eea
Since states are represented by pairs of SU(1,1) functions, the differential operators come in the form of $2\times 2$ matrices.
Finally, for SU(1,1) and a basis of $K^1$ eigenstates, we coordinatize the group as in \eq{parametrizationSU(1,1)K^1}, so that
\bea
K^1 &=& -\irm \pa_u\,\mathbbm{1}\,, \\
\clF^\pm &=& \irm\, \e^{\mp u}\left(\tanh t\, \pa_u \pm \pa_t - \frac{1}{\cosh t}\,\pa_\varphi\right) \mathbbm{1}\,, \\
J^1 &=& \big[-(\rho/2 + \irm) \sinh t - \irm\cosh t\,\pa_t\big]\, \sigma_3\,.
\eea
In the above parametrizations,  $J^3$ and $K^1$ are given by a single derivative, and the corresponding
$\SLC$ counterparts $K^3$ and $J^1$ have a particularly simple form as well.
Note that we use differential operators on the group, which is parametrized by three variables,
while Mukunda works with quotient spaces of the group, which have only two coordinates \cite{Mukunda}.

\subsection{Action on state functions}
\label{actiononstatefunctions}

Our next task is to apply the differential operators $K^3$ and $J^1$ on the state functions specified in \sec{explicitexpressions}.
This step is facilitated by the fact that, in their respective parametrizations, the operators and states for the different cases
are all of a similar form.

In each case, the selected $\SLC$ generator ($K^3$ or $J^1$) depends only on one of the three coordinates of the group, so that it acts only
on the $d$-- or $b$--function within the $D$--function. It is convenient to express this operator in terms of the variable $z$ which was used
earlier when defining the $d$-- and $b$--functions (see eqns.\ \eq{dfunctionSU(2)}, \eq{bfunctionSU(1,1)J^3}, \eq{dfunctionSU(1,1)K^1discrete}
and \eq{dfunctionSU(1,1)K^1continuous}). In fact, for all cases, the operator is essentially of the form
\be
\Oh \equiv (\rho/2 + \irm) (1 - 2z) + 2\irm z(1-z)\pa_z\,.
\ee
For SU(2), we have $K^3 = -\Oh$, 
for SU(1,1) in the $J^3$--adapted parametrization, we find $K^3 = -\Oh\, \sigma_3$, 
and for SU(1,1) in the $K^1$--adapted coordinates $J^1 = \irm \Oh\, \sigma_3$. 
The $d$-- and $b$--functions, on the other hand, are all given by linear combinations of the function $F^j_{m'm}(z)$ in eq.\ \eq{Fjm'm}.
Thus, the problem is essentially reduced to finding the action of the operator $\Oh$ on the function $F^j_{m'm}(z)$.

This action can be determined from rather lengthy manipulations of hypergeometric functions which we delegate to appendix \ref{derivationofmainequation}.
The result is that for $j \neq -1$
\be
\Oh F^j_{m'm} =
\big[\rho/2 + \irm(j+1)\big] C^{j+1}_{m'm} F^{j+1}_{m'm}
+ \frac{1}{2}\rho\, C^j_{m'm} F^j_{m'm}
+ (\rho/2 - \irm j) C^{j-1}_{m'm} F^{j-1}_{m'm}\,,
\label{mainequation}
\ee
with the coefficients given by \setlength{\jot}{0.3cm}
\bea
C^{j+1}_{m'm} &=& \frac{(j + m' + 1)(j - m + 1)}{(j+1)(2j+1)}\,,
\label{coefficients1} \\
C^j_{m'm} &=& \frac{m'm}{j(j+1)}\,,
\label{coefficients2} \\
C^{j-1}_{m'm} &=& \frac{(j - m')(j + m)}{j(2j+1)}\,.
\label{coefficients3}
\eea \setlength{\jot}{0.5cm}
This is the central equation from which the matrix elements for all cases follow.
The equation for the special value $j= -1$ arises from the limit $j\to -1$ of eq.\ \eq{mainequation}, which yields
\bea
\lim_{j\to -1} \Oh F^j_{m'm}(z) &=&
- \rho m' m\, \pa_j F^j_{m'm}(z)\big|_{j = -1} \nonumber \\
&& {}+ \frac{1}{2}\big[\rho m' m - \rho(m' - m) + 2\irm m' m\big] F^{-1}_{m'm}(z) \nonumber \\
&& - (\rho/2 + \irm)(m' + 1)(m - 1) F^{-2}_{m'm}(z)\,.
\label{mainequationlimit}
\eea

\subsection{Treatment of normalization factors}
\label{treatmentofnormalizationfactors}

Once eq.\ \eq{mainequation} is established, it remains to include the normalization factors
in the calculation in order to obtain the action of the generators on $D$--functions and states.
Since the normalization is spin dependent, the coefficients $C^{j+1}_{m'm}$
and $C^{j-1}_{m'm}$ in \eq{mainequation} have to be adjusted by additional factors that compensate
for the change from $j$ to $j+1$ or $j-1$ in the states.

We will not present the complete derivation of these factors, but comment on some of the subtleties.
In the derivation for SU(2), we can first consider the values $m' \ge m$ and $m' + m \ge 0$, in which case
the definition \eq{dfunctionSU(2)} of the $d$--function applies. The result is \eq{K^3SU(2)}.
For the other cases, we use table \ref{tabledfunction} and the property that $m$ and $m'$ appear either as
$\sqrt{m'{}^2 - j^2}\sqrt{m^2 - j^2}$ or $m m'$ in \eq{K^3SU(2)}, which are invariant under $m' \leftrightarrow m$ and $(m',m) \to (-m',-m)$.
Hence the matrix elements have the same form as in the first case.

For the discrete series of SU(1,1) in the $J^3$ basis, the calculation is analogous except for a sign factor under the square root in
$\Nt^j_{m'm}$ and a factor $\tau$ from the $\sigma_3$ of the differential operator.
The former is compensated by the fact that
$\sqrt{m'{}^2 - j^2}\sqrt{m^2 - j^2}$ switches sign as we go from SU(2) to SU(1,1) and the $\tau$ is cancelled in the $j$ term,
since $m' = \tau n / 2$.
When dealing with the continuous series of SU(1,1) in the $J^3$ basis, square roots have to be treated with particular care,
since $j$ is complex and, in general, $\sqrt{z_1}\sqrt{z_2} \neq \sqrt{z_1 z_2}$ for complex numbers $z_1$ and $z_2$.
Consider the case $m' \ge m$ for which \eq{bfunctionSU(1,1)J^3} holds.
We first note that
\be
\prod_{l = 0}^{m' - m - 1} (j + m' - l)(m - j + l) = \prod_{l = 0}^{m' - m - 1} (m - j + l)(m + j + 1 + l)
\label{rewritingproduct}
\ee 
Since $(m - j + l)^* = (m + j + 1 + l)$, this implies that
\be
\left[\prod_{l = 0}^{m' - m - 1} (j + m' - l)(m - j + l)\right]^\half = \prod_{l = 0}^{m' - m - 1} (j + m' - l)^\half (m - j + l)^\half\,.
\label{splittingsquareroot}
\ee

\renewcommand{\arraystretch}{2}
Furthermore, one can show that for all $j = -\frac{1}{2} + \irm s$, $s > 0$, and $m$,
\bea
|\arg(m + j + 1) + \arg(m - j - 1)| &<& \pi\,, \\
|\arg(m + j) + \arg(m - j)| &<& \pi\,,
\eea 
and
\bea
|\arg(m + j + 1) - \arg(m - j - 1)|
&\!\!&\left\{
\begin{array}{ll}
< \pi\,, & m \ge \frac{1}{2}\,, \\
= \pi\,, & m = 0\,,
\end{array}
\right. \\
|\arg(m - j) - \arg(m + j)|
&\!\!&
\left\{
\begin{array}{ll}
< \pi\,, & m \ge \frac{1}{2}\,, \\
= -\pi\,, & m = 0\,.
\end{array}
\right.
\eea
This allows us to write
\be
\sqrt{m + j + 1}\sqrt{m - j - 1} = \sqrt{m^2 - (j + 1)^2}\,,\qquad
\sqrt{m + j}\sqrt{m - j} = \sqrt{m^2 - j^2}\,,
\label{multiplysquareroot}
\ee
and for $m \ge \frac{1}{2}$ $(m = 0)$,
\be
\frac{\sqrt{m + j + 1}}{\sqrt{m - j - 1}} = \sqrt{\frac{m + j + 1}{m - j - 1}}\,,\qquad
\frac{\sqrt{m - j}}{\sqrt{m + j}} = \pm\sqrt{\frac{m - j}{m + j}}\,.
\label{dividesquareroot}
\ee
With the help of equations \eq{splittingsquareroot} and \eq{multiplysquareroot} one arrives at
the final formula \eq{K^3J^3states}. The case $m' < m$ follows from table \ref{tabledfunction}.

When coming to the discrete series in the $K^1$ basis, we observe that for $m \ge -j$
\be
S^j_m = \sqrt{m^2 - (j+1)^2}\,S^{j+1}_m\,,\qquad
S^j_m = \frac{S^{j-1}_m }{\sqrt{m^2 - j^2}}\,.
\ee 
We first derive eq.\ \eq{J^1discrete} for $\tau = 1$, for which $m = n/2 \ge -j$.
Then, the case $\tau = -1$ is inferred from inspection of eq.\ \eq{dfunctionK^1discrete_mnegative}:
$m$ goes to $-m$ in the coefficients and the change $t\to -t$ amounts to a sign for all three terms;
the substitution $m\to -m$ has no net effect, however, since also $m = \tau n/2$, and the sign from $t\to -t$ is cancelled by the
sign from $\sigma_3$ in $J^1$. Thus, the coefficients have the same form as for $\tau = 1$.

In the case of the continuous series, we use $\Gamma(m - j)\Gamma(m + j + 1) \ge 0$
together with eqns.\ \eq{multiplysquareroot} and \eq{dividesquareroot} to obtain
\be
S^j_m = \epsilon_{j+1} \sqrt{m^2 - (j+1)^2}\,S^{j+1}_m\,,\qquad
S^j_m = \epsilon_j \frac{S^{j-1}_m }{\sqrt{m^2 - j^2}}\,.
\label{Sj}
\ee 
where $\epsilon_{j+1} = \pm 1$ for $m \ge 0$ $(m \le -\frac{1}{2})$ and $\epsilon_j = \pm 1$ for $m \ge \frac{1}{2}$ $(m \le 0)$.
To derive the matrix elements we need furthermore the identities
\be
T^j_{m\lambda\sigma} = \irm\, \frac{(-j + \irm\lambda - 1)}{2(-m - j - 1)}\, T^{j+1}_{m\lambda\sigma'}\,,\qquad
T^j_{m\lambda\sigma} = -\irm\, \frac{2(-m - j)}{-j + \irm\lambda}\, T^{j-1}_{m\lambda\sigma'}\,.
\ee 
Given eq.\ \eq{J^1continuous} for $\tau = 1$, we deduce the $\tau = -1$ component by noting the following.
The sign from $\sigma_3$ in $J^1$ produces an overall sign for all three terms. When $m = \tau n/2 < 0$, we get another
sign for the $j$ term, and a sign for the $j+1$ and $j-1$ term from relation \eq{Sj}.
Therefore, the coefficients are identical to those for $\tau = 1$. On the other hand, if $n = 0$ and hence $m = \tau n/2 = 0$,
eq.\ \eq{Sj} gives only a sign for the $j-1$ term. In this case, there remains a sign change for the $j+1$ term.

\section{Summary and discussion}
\label{summary}

In this paper, we have determined the matrix elements of generators in SU(1,1)
decompositions of unitary irreducible representations of $\SLC$.
By extending and building on previous work by Mukunda \cite{Mukunda},
we derived these matrix elements for both the discrete basis diagonal
in $J^3$ and the continuous basis of $K^1$ eigenstates, and in each case for
the discrete and continuous series.
By identifying the common structure of differential operators and states across
different bases, the problem was reduced to one main equation. Basis--specific differences
appeared in the treatment of normalization factors.

As explained in the introduction, unitary representations of $\SLC$
and its states are central elements in the spin foam approach to quantum gravity.
It was shown by us in ref.\ \cite{CHtimelike} that coherent states of the SU(1,1) reduction
represent quantum states of timelike 2--cells\footnote{To be precise, states in the $K^1$ basis
and continuous series correspond to timlike 2--cells, while states in the $J^3$ basis and discrete
series implement spacelike 2--cells.}.
For this, we used the matrix elements of these states, anticipating the proof of the present paper.
It is also likely that these matrix elements will be relevant for future calculations
in spin foam and loop quantum gravity.

Finally, this paper could be useful for anybody interested in
the reduction of $\SLC$ representations, since it collects some of the
know--how that is dispersed over references from more than 40 years ago.

\begin{acknowledgments}
JH would like to thank the Natural Sciences and Engineering Research Council of Canada (NSERC) for his post graduate scholarship.
Research at Perimeter Institute is supported by the Government of Canada through Industry Canada and by the Province of Ontario through the Ministry of Research \& Innovation.
\end{acknowledgments}

\begin{appendix}

\section{Parametrization of $\SLC$, SU(2) and SU(1,1)}
\label{parametrization}

In this section, we state our conventions for parametrizations and measures on $\SLC$, SU(2) and SU(1,1).
The one--parameter subgroups of $\SLC$ are parametrized as follows:
\[
\begin{array}{l@{\qquad}l}
J_1 = \frac{1}{2}\sigma_1\,, & a_1(\psi) = \e^{\irm\psi J_1} = \twomatrix{\cos(\psi/2)}{\irm\sin(\psi/2)}{\irm\sin(\psi/2)}{\cos(\psi/2)} \\
\\
J_2 = \frac{1}{2}\sigma_2\,, & a_2(\theta) = \e^{\irm\theta J_2} = \twomatrix{\cos(\theta/2)}{\sin(\theta/2)}{-\sin(\theta/2)}{\cos(\theta/2)} \\
\\
J_3 = \frac{1}{2}\sigma_3\,, & a_3(\varphi) = \e^{\irm\varphi J_3} = \twomatrix{\e^{\irm\varphi/2}}{0}{0}{\e^{-\irm\varphi/2}} \\
\\
K_1 = \frac{\irm}{2}\sigma_1\,, & b_1(u) = \e^{\irm u K_1} = \twomatrix{\cosh(u/2)}{-\sinh(u/2)}{-\sinh(u/2)}{\cosh(u/2)} \\
\\
K_2 = \frac{\irm}{2}\sigma_2\,, & b_2(t) = \e^{\irm t K_2} = \twomatrix{\cosh(t/2)}{\irm\sinh(t/2)}{-\irm\sinh(t/2)}{\cosh(t/2)} \\
\\
K_3 = \frac{\irm}{2}\sigma_3\,, & b_3(\delta) = \e^{\irm \delta K_3} = \twomatrix{\e^{-\delta/2}}{0}{0}{\e^{\delta/2}} \\
\end{array}
\] \setlength{\jot}{0.3cm}
For elements $u$ of SU(2) we use the parametrization
\be
u = \e^{\irm\psi J^3} \e^{\irm\theta J^2} \e^{\irm\varphi J^3}\,,\qquad 0 \le \psi < 4\pi\,,\quad 0 \le \theta < \pi\,,\quad -\pi \le \varphi < \pi\,.
\label{parametrizationSU(2)}
\ee 
In these coordinates, the normalized Haar measure takes the form
\be
\d u = \frac{1}{(4\pi)^2} \sin\theta\, \d\psi\, \d\theta\, \d\varphi\,.
\ee
For SU(1,1) elements $v$ we adopt two kinds of parametrizations.
When using a $J^3$ basis, we employ
\be
v = \e^{\irm\psi J^3} \e^{\irm t K^2} \e^{\irm\varphi J^3}\,,\qquad 0 \le \psi < 4\pi\,,\quad 0 \le t < \infty\,,\quad -\pi \le \varphi < \pi\,,
\label{parametrizationSU(1,1)J^3}
\ee 
together with the measure
\be
\d v = \frac{1}{(4\pi)^2} \sinh t\, \d\psi\, \d t\, \d\varphi\,.
\label{measureSU(1,1)}
\ee
In the case of $K^1$ eigenstates, we use instead the following parametrization given in \cite{Lindblad},
\be
v = \e^{\irm\varphi J^3} \e^{\irm t K^2} \e^{\irm u K^1}\,,\qquad 0 \le \varphi < 4\pi\,,\quad 0 \le t, u < \infty\,,
\label{parametrizationSU(1,1)K^1}
\ee 
for which the measure \eq{measureSU(1,1)} reads
\be
\d v = \frac{1}{(4\pi)^2} \cosh t\, \d\varphi\, \d t\, \d u\,.
\ee

\section{Representation functions of SU(1,1)}
\label{representationfunctionsofSU(1,1)}

\renewcommand{\arraystretch}{1.5}
We verify below that the definition of the $D$--functions in eqns.\ \eq{DfunctionSU(1,1)J^3} and \eq{bfunctionSU(1,1)J^3}
is equivalent to the original expressions derived by Bargmann (\cite{Bargmann}, see also \cite{HolmesBiedenharn}). 
Let us write a general SU(1,1) element as
\be
v = \twomatrix{\alpha}{\beta}{\beta^*}{\alpha^*}\,,\qquad \alpha, \beta\in \bC\,.
\ee \renewcommand{\arraystretch}{2}
For the discrete series, Bargmann gives the following definitions of the $D$--functions.

For $m', m \ge -j$,
\bean
\lefteqn{D^j_{m'm}(v)} \\
&&=
\left\{
\begin{array}{ll}
\Theta_{m'm}\, {\alpha^*}^{-(m'+m)} \beta^{m'-m} F(-j-m,j-m+1,m'-m+1,-|\beta|^2)\,, & m' \ge m \\
(-1)^{m-m'} \Theta_{m'm}\, {\alpha^*}^{-(m'+m)} {\beta^*}^{m-m'} F(-j-m',j-m'+1,m-m'+1,-|\beta|^2)\,, & m' < m
\end{array}
\right. \\
&&
\Theta_{m'm} = \frac{1}{(m'-m)!} \left[\frac{(m'+j)! (m'-j-1)!}{(m+j)! (m-j-1)!}\right]^\half
\eean
For $m', m \le j$,
\bean
\lefteqn{D^j_{m'm}(v)} \\
&&=
\left\{
\begin{array}{ll}
\Theta_{m'm}\, \alpha^{m'+m} \beta^{m'-m} F(-j+m',j+m'+1,m'-m+1,-|\beta|^2)\,, & m' \ge m \,,\\
(-1)^{m-m'} \Theta_{mm'}\, \alpha^{m'+m} {\beta^*}^{m-m'} F(-j+m,j+m+1,m-m'+1,-|\beta|^2)\,, & m' < m\,,
\end{array}
\right. \\
&& \Theta_{m'm} = \frac{1}{(m'-m)!} \left[\frac{(-m+j)! (-m-j-1)!}{(-m'+j)! (-m'-j-1)!}\right]^\half
\eean
By using the parametrization \eq{parametrizationSU(1,1)J^3},
\be
\alpha = \e^{\irm(\psi+\varphi)/2} \cosh(t/2)\,,\quad \beta = \irm\, \e^{\irm(\psi-\varphi)/2} \sinh(t/2)\,,
\ee
and identity 2.9 (2) from Bateman/Erdelyi \cite{Bateman},
\be
F(a,b,c;z) = (1-z)^{c-a-b} F(c-b,c-a,c;z)\,,
\ee \renewcommand{\arraystretch}{2}
one can rewrite these $D$--functions in the form stated in eqns.\ \eq{DfunctionSU(1,1)J^3} and \eq{bfunctionSU(1,1)J^3} of the main text. 
In the case of the continuous series, Bargmann defines
\bean
\lefteqn{D^j_{m'm}(v)} \\
&&=
\left\{
\begin{array}{ll}
\Theta_{m'm}\, \alpha^{m'+m} \beta^{m'-m} F(j+m'+1,-j+m',m'-m+1,-|\beta|^2)\,, & m' \ge m\,, \\
(-1)^{m-m'} \Theta_{mm'}\, \alpha^{m'+m} {\beta^*}^{m-m'} F(j+m+1,-j+m,m-m'+1,-|\beta|^2)\,, & m' < m\,,
\end{array}
\right. \\
&& \Theta_{m'm} = \frac{1}{(m'-m)!}\prod_{k=1}^{m'-m}\left[\frac{1}{4} + s^2 + (m+k)(m+k-1)\right]^\half \,,\qquad m' \ge m\,.
\eean
By eq.\ \eq{rewritingproduct} this is equivalent to eqns.\ \eq{DfunctionSU(1,1)J^3} and \eq{bfunctionSU(1,1)J^3}.

\section{Derivation of main equation}
\label{derivationofmainequation}


In this section, we derive the main equation for the determination of the matrix elements (eq.\ \eq{mainequation}).
The proof consists of two parts, corresponding to the multiplicative and derivative part of the operator $\Oh$
respectively.
An explicit treatment of the multiplicative term has been given in the appendix of \cite{Mukunda}, so we will only
quote the result:
\be
(1 - 2z) F^j_{m'm}(z) = C^{j+1}_{m'm} F^{j+1}_{m'm}(z) + C^j_{m'm} F^j_{m'm}(z) + C^{j-1}_{m'm} F^{j-1}_{m'm}(z)
\label{A9fromMukunda}
\ee
The coefficients are the ones defined in eqns. \eq{coefficients1}--\eq{coefficients3}.
The statement of the full equation \eq{mainequation} and the proof for the derivative part
were omitted in \cite{Mukunda}. We will provide this derivation now. In the following, the hypergeometric function
$_2F_1$ is abbreviated by $F$.

The starting point is identity 2.8 (27) in Bateman/Erdelyi \cite{Bateman}:
\be
(c-1)\, z^{c-2} (1-z)^{a+b-c-1} F(a-1,b-1;c-1;z)
= \diff{}{z} \left[z^{c-1} (1-z)^{a+b-c}\, F(a,b;c;z)\right]
\ee
Setting
\be
a = -j+m'\,,\qquad b = j+m'+1\,,\qquad c = m'-m+1\,,
\label{valuesofabc}
\ee
one can write the function $F^j_{m'm}$ as
\be
F^j_{m'm}(z) = \frac{1}{(c-1)!}\, z^{(c-1)/2} \left(1-z\right)^{(a+b-c)/2} F(a,b;c;z)\,.
\ee
Then, eq.\ 2.8 (27) implies
\bea
\lefteqn{z^{-(c-1)/2} \left(1-z\right)^{-(a+b-c)/2}\diff{}{z}\left[z^{(c-1)/2} \left(1-z\right)^{(a+b-c)/2} F^j_{m'm}(z)\right]} \\
&&=
(c-1)\,z^{-1}\left(1-z\right)^{-1} z^{(c-1)/2} \left(1-z\right)^{(a+b-c)/2} F(a-1,b-1;c-1;z) / (c-1)!\,.
\eea
From here a brief calculation leads to
\bea
z(1-z)\diff{}{z} F^j_{m'm}(z)
&=&
\frac{1}{2} m\, F^j_{m'm}(z) - \frac{1}{2} m'\, (1-2z) F^j_{m'm}(z) \nonumber \\
&&
\hspace{-2.5cm}{}+ (m'-m)\,\frac{1}{(m'-m)!} \left(1-z\right)^{(m'+m)/2} z^{(m'-m)/2} F(a-1,b-1;c-1;z)\,.
\label{beforedealingwithFa-1b-1c-1}
\eea
The aim is to decompose the right--hand side into functions $F(a-1,b+1;c;z)$, $F(a,b;c;z)$ and $F(a+1,b-1;c;z)$
which will give the $j+1$, $j$ and $j-1$ term of the final equation.
To decompose $F(a-1,b-1;c-1;z)$ suitably we use several identities from Bateman/Erdelyi.
From 2.8 (35) it follows that
\be
F(a-1,b-1;c-1;z) = \frac{c-a}{c-1} F(a-1,b-1;c;z) + \frac{a-1}{c-1} F(a,b-1;c;z)\,.
\label{1}
\ee
Identity 2.8 (33) implies that
\be
F(a-1,b-1;c;z) = \frac{c-a-b+1}{c-b} F(a-1,b;c;z) + \frac{a-1}{c-b} (1-z) F(a,b;c;z)\,.
\label{2}
\ee
From 2.8 (32) we get
\be
F(a-1,b;c;z) = \frac{b}{b-a+1} F(a-1,b+1;c;z) - \frac{a-1}{b-a+1} F(a,b;c;z)\,.
\label{3}
\ee
Likewise, one obtains
\be
F(a,b-1;c;z) = \frac{b-1}{b-a-1} F(a,b;c;z) - \frac{a}{b-a-1} F(a+1,b-1;c;z)\,.
\label{2p}
\ee
We start from \eq{1} and plug in \eq{2} and \eq{2p}, giving us
\bea
\lefteqn{F(a-1,b-1;c-1;z)} \nonumber \\
&&=
\frac{c-a}{c-1} \left[\frac{c-a-b+1}{c-b} F(a-1,b;c;z) + \frac{a-1}{c-b} (1-z) F(a,b;c;z)\right] \nonumber \\
&& {}+ \frac{a-1}{c-1} \left[\frac{b-1}{b-a-1} F(a,b;c;z) - \frac{a}{b-a-1} F(a+1,b-1;c;z)\right]\,.
\eea
By inserting \eq{3}, we arrive at
\bea
\lefteqn{F(a-1,b-1;c-1;z)} \nonumber \\
&&=
\frac{(c-a)(c-a-b+1)b}{(c-1)(c-b)(b-a+1)}\, F(a-1,b+1;c;z) \nonumber \\
&&
{}- \frac{(a-1)a}{(c-1)(b-a-1)}\, F(a+1,b-1;c;z) \nonumber \\
&&
{}+ \frac{a-1}{c-1}\left[-\frac{(c-a-b+1)(c-a)}{(c-b)(b-a+1)} + \frac{c-a}{2(c-b)} + \frac{b-1}{b-a-1}\right] F(a,b;c;z)
\nonumber \\
&&
{}+ \frac{(c-a)(a-1)}{2(c-1)(c-b)}\, (1-2z) F(a,b;c;z)\,.
\eea
When inserting the values \eq{valuesofabc} this assumes the form
\bea
\lefteqn{F(a-1,b-1;c-1;z)} \\
&&=
-\frac{(j - m + 1)(-m' - m + 1)(j+m'+1)}{2(m' - m)(j+m)(j+1)}\, F(a-1,b+1;c;z) \\
&&
{}- \frac{(-j + m' - 1)(-j+m')}{2(m' - m)j}\, F(a+1,b-1;c;z) \\
&&
{}+ \frac{-j + m' - 1}{2(m' - m)}\left[\frac{(-m' - m + 1)(j - m + 1)}{(j+m)(j+1)} - \frac{j - m + 1}{j+m} + \frac{j+m'}{j}\right] F(a,b;c;z)
\nonumber \\
&&
{}- \frac{(j - m + 1)(-j + m' - 1)}{2(m' - m)(j+m)}\, (1-2z) F(a,b;c;z)\,.
\eea
Having obtained a formula for $F(a-1,b-1;c-1;z)$, we can proceed with eq.\ \eq{beforedealingwithFa-1b-1c-1}, namely,
\bea
z(1-z)\pa_z F^j_{m'm}(z)
&=&
-\frac{(j - m + 1)(-m' - m + 1)(j+m'+1)}{2(j+m)(j+1)}\, F^{j+1}_{m'm}(z) \nonumber \\
&&
{}- \frac{(-j + m' - 1)(-j+m')}{2j}\, F^{j-1}_{m'm}(z) \nonumber \\
&&
\hspace{-2cm}{}+ \frac{1}{2}\left\{m + (-j + m' - 1)\left[\frac{(-m' - m + 1)(j - m + 1)}{(j+m)(j+1)} - \frac{j - m + 1}{j+m} + \frac{j+m'}{j}\right]\right\} F^j_{m'm}(z)
\nonumber \\
&&
{}+ \frac{1}{2}\left[-m' - \frac{(j - m + 1)(-j + m' - 1)}{j+m}\right] (1-2z) F^j_{m'm}(z)\,.
\label{beforeplugginginA9}
\eea
The final step is to insert $(1-2z) F^j_{m'm}(z)$ from eq.\ \eq{A9fromMukunda}, which results in
\bean
&& z(1-z)\pa_z F^j_{m'm}(z) \\
&=&
\left\{-\frac{(j - m + 1)(-m' - m + 1)(j+m'+1)}{2(j+m)(j+1)}\right. \\
&&
\hspace{0.2cm}\left.{}+ \frac{1}{2}\left[-m' - \frac{(j - m + 1)(-j + m' - 1)}{j+m}\right] \frac{(j+m'+1)(j-m+1)}{(j+1)(2j+1)}\right\} F^{j+1}_{m'm}(z) \\
&&
\hspace{-0.5cm}{}+ \left\{- \frac{(-j + m' - 1)(-j+m')}{2j}
+ \frac{1}{2}\left[-m' - \frac{(j - m + 1)(-j + m' - 1)}{j+m}\right] \frac{(j-m')(j+m)}{j(2j+1)}\right\} F^{j-1}_{m'm}(z) \\
&&
{}+ \frac{1}{2}\left\{m + (-j + m' - 1)\left[\frac{(-m' - m + 1)(j - m + 1)}{(j+m)(j+1)} - \frac{j - m + 1}{j+m} + \frac{j+m'}{j}\right]\right. \\
&&
\hspace{1.2cm}{}+ \left.\left[-m' - \frac{(j - m + 1)(-j + m' - 1)}{j+m}\right] \frac{m' m}{j(j+1)}\right\} F^j_{m'm}(z)\,.
\eean
Simplification yields
\be
z(1-z)\pa_z F^j_{m'm}(z) = \frac{1}{2}j\,C^{j+1}_{m'm} F^{j+1}_{m'm}(z) - \frac{1}{2}\,C^j_{m'm} F^j_{m'm}(z) - \frac{1}{2}(j+1)\,C^{j-1}_{m'm} F^{j-1}_{m'm}(z)\,.
\ee
By combining this with the multiplicative part \eq{A9fromMukunda} we arrive at eq.\ \eq{mainequation} in the main part of the article.
\qed

\end{appendix}


\begin{thebibliography}{99}

\bibitem{Rovellibook}
C.~Rovelli, ``Quantum Gravity'',
Cambridge University Press, Camrbidge (2004).

\bibitem{Thiemannbook}
T.~Thiemann,
``Modern canonical quantum general relativity'',
Cambridge University Press, Cambridge (2007).

\bibitem{ReisenbergerRovelliFeynman1}
M.~Reisenberger, C.~Rovelli,
\textit{Spin foams as Feynman diagrams},
[arXiv:gr-qc/0002083].

\bibitem{ReisenbergerRovelliFeynman2}
M.P.~Reisenberger and C.~Rovelli,
\textit{Spacetime as a Feynman diagram: The connection formulation},
Class.Quant.Grav.\ {\bf 18} 121 (2001),
[arXiv:gr-qc/0002095].

\bibitem{Oritigroupfieldtheoryapproach}
D.~Oriti, {\it The group field theory approach to quantum gravity},
[arXiv:gr-qc/0607032].

\bibitem{BarrettCrane}
J.W.~Barrett, L.~Crane, \textit{Relativistic spin networks and
quantum gravity}, J.Math.Phys.\  {\bf 39}, 3296 (1998),
[arXiv:gr-qc/9709028].

\bibitem{EPRL}
J.~Engle, E.~Livine, R.~Pereira and C.~Rovelli, {\it LQG vertex with
finite Immirzi parameter}, Nucl.Phys.\ \textbf{B799}, 136 (2008),
[arXiv:0711.0146 [gr-qc]].

\bibitem{FK}
L.~Freidel, K.~Krasnov, {\it A New Spin Foam Model for 4d
Gravity}, Class.Quant.Grav.\ \textbf{25}, 125018 (2008),
[arXiv:0708.1595 [gr-qc]].

\bibitem{Livine_Consistently}
E.R.~Livine, S.~Speziale, {\it Consistently Solving the
Simplicity Constraints for Spinfoam Quantum Gravity},
Europhys.Lett.\ \textbf{81}, 50004 (2008), [arXiv:0708.1915
[gr-qc]].

\bibitem{Livine_New}
E.R.~Livine, S.~Speziale, {\it A new spinfoam vertex for quantum
gravity}, Phys.Rev.\ \textbf{76D}, 084028 (2007), [arXiv:0705.0674
[gr-qc]]

\bibitem{CFpathrep}
F.~Conrady, L.~Freidel,
\textit{Path integral representation of spin foam models of 4d gravity},
Class.Quant.Grav.\ {\bf 25}, 245010 (2008),
[arXiv:0806.4640 [gr-qc]].

\bibitem{CFsemiclassical}
F.~Conrady, L.~Freidel,
\textit{On the semiclassical limit of 4d spin foam models},
Phys.Rev.\ {\bf D78}, 104023 (2008),
[arXiv:0809.2280 [gr-qc]].

\bibitem{BarrettasymptoticsEuclidean}
J.W.~Barrett, R.J.~Dowdall, W.J.~Fairbairn, H.~Gomes, F.~Hellmann,
{\it Asymptotic analysis of the EPRL four-simplex
amplitude}, J.Math.Phys.\ \textbf{50}, 112504 (2009),
[arXiv:0902.1170 [gr-qc]].

\bibitem{BarrettasymptoticsLorentzian}
J.W.~Barrett, R.J.~Dowdall, W.J.~Fairbairn, F.~Hellmann, R.~Pereira, {\it Lorentzian spin foam amplitudes: graphical calculus
and asymptotics}, [arXiv:0907.2440 [gr-qc]].

\bibitem{CFquantumgeometry}
F.~Conrady, L.~Freidel,
\textit{Quantum geometry from phase space reduction},
J.Math.Phys.\ {\bf 50}, 123510 (2009),
[arXiv:0902.0351 [gr-qc]].

\bibitem{CHtimelike}
F.~Conrady, J.~Hnybida, {\it A spin foam model for general Lorentzian 4--geometries},
[arXiv:1002.1959 [gr-qc]].

\bibitem{Ccoherent}
F.~Conrady,
\textit{Spin foams with timelike surfaces},
Class.Quant.Grav.\ {\bf 27}, 155014 (2010), [arXiv:1003.5652 [gr-qc]].

\bibitem{LindbladNagel}
G.~Lindblad, B.~Nagel, {\it Continuous bases for unitary
irreducible representations of SU(1,1)}, Ann. De L'I.H.P., Sec. A.\
\textbf{13}, 27-56 (1970).

\bibitem{MukundaO21inO11basis}
N.~Mukunda, \textit{Unitary representations of the group O(2,1) in an O(1,1) basis}, J.Math.Phys.\ \textbf{8}, 2210 (1967).

\bibitem{Naimark}
M.A.~ Naimark, ``Linear representations of the Lorentz group'', Pergamon Press, 1964.

\bibitem{HarishChandra}
Harish--Chandra, \textit{Infinite irreducible representations of the Lorentz group}, Proc.Roy.Soc.\ \textbf{A189}, 372 (1947).

\bibitem{SciarrinoToller}
A.~Sciarrino, M.~Toller, \textit{Decomposition of the Unitary Irreducible Representations of the Group SL(2C) Restricted to the Subgroup SU(1,1)},
J.Math.Phys. \textbf{8}, 1252-1265 (1967).

\bibitem{Delbourgo}
R.~Delbourgo, K.~Koller, P.~Mahanta,
\textit{On transformations between SL(2,C) representations}.
Nuovo Cim.\  \textbf{52A}, 1254 (1967).

\bibitem{Ruhl}
W.~Ruhl, ``Lorentz group and harmonic analysis'', W.A.~Benjamin, New York, (1970).

\bibitem{Carmeli}
M.~Carmeli, ``Group theory and general relativity'', McGraw-Hill,
New York (1977).

\bibitem{Strom1967vol33}
S.~Strom, \textit{A note on the matrix elements of a unitary representation of the homogeneous Lorentz group},
Arkiv Fysik \textbf{33}, 465 (1967).

\bibitem{DucVanHieu}
D.V.~Duc, N.~Van Hieu, \textit{On the theory of unitary representations of the SL(2,C) group},
Ann.Inst.~Henri Poincare, \textbf{6}, 17--37 (1967).

\bibitem{Mukunda}
N.~Mukunda, {\it Unitary Representations of the Homogeneous Lorentz
Group in an O(2,1) Basis}, J.Math.Phys.\ \textbf{9}, 50 (1968).

\bibitem{AndrewsGunson}
M.~Andrews, J.~Gunson,
\textit{Complex angular momenta and many--particle states. I.~Properties of local representations of the rotation group},
J.Math.Phys. \textbf{5}, 1391 (1964).

\bibitem{Bargmann}
V.~Bargmann,
\textit{Irreducible unitary representations of the Lorentz group},
Annals Math.\  \textbf{48}, 568 (1947).

\bibitem{Lindblad}
G.~Lindblad, {\it Eigenfunction expansions associated with unitary
irreducible representations of SU(1,1)}, Phys.Scripta \textbf{1},
201 (1970).

\bibitem{Tung}
W.K.~Tung,
``Group Theory In Physics'', Singapore, World Scientific (1985).

\bibitem{Ui}
H.~Ui,
\textit{Clebsch-Gordan formulas of the SU(1,1) group},
Prog.Theor.Phys.\ \textbf{44}, 689 (1970).

\bibitem{Byrd}
M.S.~Byrd,
\textit{Differential geometry on SU(3) with applications to three state systems},
J.Math.Phys.\ \textbf{39}, 6125–-6136 (1998).

\bibitem{HolmesBiedenharn}
W.J.~Holman III, L.C.~Biedenharn Jr.,
\textit{Complex angular momenta and the groups SU(1, 1) and SU(2)},
Ann.Phys.\ \textbf{39}, 1--42 (1966).

\bibitem{Bateman}
H.~Bateman, A.~Erdelyi, ``Higher Transcendental Functions'',
Vol.\ 1, McGraw-Hill, 1953.

\end{thebibliography}
\end{document}